\documentclass[conference]{IEEEtran}
\IEEEoverridecommandlockouts
\usepackage{cite}
\usepackage{amsmath,amssymb,amsfonts}

\usepackage{booktabs} 
\usepackage{multirow} 
\usepackage{pgfplots} 
\pgfplotsset{compat=1.18}
\usepackage{algorithmic}
\usepackage{hyperref}
\usepackage{graphicx}
\usepackage{hyperref}
\usepackage{subcaption}
\usepackage{textcomp}
\usepackage{xcolor}
\usepackage{seqsplit}
\usepackage{tikz}
\usepackage{tabularx}
\usepackage{array}
\usepackage{listings}
\usepackage{adjustbox}
\usepackage{multirow}
\usepackage{pgfplots}
\usepackage{comment}
\usepackage{makecell}
\usepackage{tikz}
\usepackage{titlesec}

\titlespacing*{\section}{0pt}{1.2ex plus 0.5ex minus 0.2ex}{0.8ex}
\titlespacing*{\subsection}{0pt}{1.0ex plus 0.4ex minus 0.2ex}{0.6ex}
\usetikzlibrary{arrows.meta,positioning,fit,shadows.blur}

\pgfplotsset{width=10cm,compat=1.16}
\newcolumntype{M}[1]{>{\centering\arraybackslash}m{#1}}
\newcolumntype{P}[1]{>{\centering\arraybackslash}p{#1}}
\newcolumntype{C}[1]{>{\centering\let\newline\\\arraybackslash\hspace{0pt}}m{#1}}
\usetikzlibrary{shapes.geometric,shapes.symbols,fit,positioning,shadows, arrows.meta, backgrounds}
\def\BibTeX{{\rm B\kern-.05em{\sc i\kern-.025em b}\kern-.08em
		T\kern-.1667em\lower.7ex\hbox{E}\kern-.125emX}}
\begin{document}
	\makeatletter
	\newcommand{\linebreakand}{%
	\end{@IEEEauthorhalign}
	\hfill\mbox{}\par
	\mbox{}\hfill\begin{@IEEEauthorhalign}
	}
	\makeatother
	\title{KG-First, LLM-Fallback: A Hybrid Microservice for Grounded Skill Search and Explanation
	
	\vspace{-0.3cm}		
	}
	
	\author{
			\IEEEauthorblockN{Ngoc Luyen Le\textsuperscript{$\S$\ddag}\\\textit{ngoc-luyen.le@hds.utc.fr}}
			\and
			\IEEEauthorblockN{Marie-Hélène Abel\textsuperscript{\ddag}\\\textit{marie-helene.abel@hds.utc.fr}}\and
			\IEEEauthorblockN{Bertrand Laforge\textsuperscript{$\ast$}\\\textit{laforge@lpnhe.in2p3.fr}
					}
			\linebreakand
			
			\IEEEauthorblockA{
				\textsuperscript{$\S$} Gamaizer.ia, 93340 Le Raincy, France
			}
			\IEEEauthorblockA{
			\textsuperscript{\ddag} Université de technologie de Compiègne, CNRS, Heudiasyc (Heuristics and Diagnosis of Complex Systems), \\CS 60319 - 60203 Compiègne Cedex, France
			}
			\IEEEauthorblockA{
				\textsuperscript{$\ast$} Sorbonne Université, CNRS UMR 7585, LPMHE (Laboratoire de Physique Nucléaire et des Hautes Énergies), \\75252 Paris cedex 05, France
			}
			\vspace{-1cm}
	}

	\maketitle

	\begin{abstract}

Authoritative competency frameworks such as ESCO, ROME, and O*NET are essential for aligning education with labor market needs, yet their technical complexity and structural heterogeneity hinder practical adoption by educators. This paper introduces \emph{SkillGraph-Service}, an interoperable microservice designed to bridge this gap by unifying these resources into a provenance-preserving Knowledge Graph (KG). Adopting a \textit{KG-first, LLM-fallback} architecture, the system combines symbolic rigor with sub-symbolic flexibility. It implements a lightweight hybrid retrieval engine (fusing SQLite FTS5 and HNSW vector search) to handle the vocabulary mismatch in educator queries, and utilizes Large Language Models (LLMs) strictly for constrained ranking and audience-aware explanation. Empirical evaluation on a multilingual dataset reveals that the proposed hybrid strategy achieves superior retrieval effectiveness (nDCG@5$>$0.94) with sub-200\,ms latency, rendering computationally expensive cross-encoder re-ranking may be unnecessary for this domain. Furthermore, an analysis of generated explanations highlights a trade-off between fluency and faithfulness: while JSON-constrained LLMs ensure high citation precision, deterministic templates remain the most reliable method for maximizing evidence coverage. The resulting architecture offers a practical, scalable, and auditable solution for integrating complex skill data into digital learning ecosystems.
	\end{abstract}
	

\vspace{-0.2cm}
\section{Introduction}
Competency-based education  requires that learning activities, assessments, and outcomes be explicitly aligned with well-defined skills drawn from authoritative frameworks. In practice, however, widely used corpora such as ESCO\footnote{ESCO: European Skills, Competences, Qualifications and Occupations, \href{https://ec.europa.eu/esco}{https://ec.europa.eu/esco}}, ROME\footnote{ROME : Répertoire Opérationnel des Métiers et des Emplois, \href{https://www.francetravail.org}{https://www.francetravail.org}}, and O*NET\footnote{Occupational Information Network, \href{https://www.onetcenter.org/}{https://www.onetcenter.org/}} are heterogeneous in structure, vary in granularity and multilingual coverage, and expose technical interfaces that are difficult for non-experts to exploit~\cite{miles2009skos,sparrql11}. As a result, educators and instructional designers often rely on ad hoc keyword search and manual tagging. This limits the discoverability of relevant skills, obscures prerequisite and decomposition structures, and hinders the design of transparent, portable learning pathways.

The methodological challenge is thus to provide a lightweight, interoperable service that makes skills from multiple frameworks discoverable, interpretable, and actionable -- while strictly preserving provenance. Concretely, given a natural-language query $q$, the system should return: (i) top-$k$ relevant skills across frameworks; (ii) structural context (prerequisites, sub-skills, and granularity); and (iii) a concise, audience-aware explanation. Crucially, to support trustworthy curricular design, the solution must be grounded: it must cite evidence via stable identifiers and avoid hallucination risks typical of unconstrained generative AI.

In this work, we present \textit{``KG-First, LLM-Fallback: A Hybrid Microservice for Grounded Skill Search and Explanation''}. This architectural approach combines symbolic retrieval over a unified, provenance-aware representation of ESCO, ROME, and O*NET with the tightly controlled capabilities of Large Language Models (LLMs). The \textit{``KG-First''} component handles educator-style queries via hybrid retrieval (lexical BM25 plus sentence-level embeddings), ensuring that all search results are directly anchored in authoritative standards~\cite{robertson2009probabilistic,reimers2019sentence}. The \textit{``LLM-Fallback''} component is invoked strictly for augmentation tasks where symbolic methods fall short -- specifically, constrained query reformulation, ranking within closed candidate sets~\cite{luyen2025automated,le2025well}, and generating audience-aware explanations. These generations are strictly constrained (via templates or JSON schemas) to remain grounded in the provided graph facts~\cite{da2025capability}.

We investigate two primary research questions: whether this hybrid retrieval strategy improves educator-style skill search over unimodal baselines (RQ1); and whether template-based and JSON-constrained explanations yield higher clarity and faithfulness (evidence-linked) compared to free-form generation under practical latency constraints (RQ2). 
We implement \emph{SkillGraph-Service}, a prototype exposing these capabilities via a minimal REST API, and demonstrate its integration readiness through an LMS-oriented workflow scenario with initial empirical results.

The remainder of this paper is organized as follows: Section~\ref{sec:related} reviews related work in semantic interoperability and neural retrieval. Section~\ref{sec:architecture} details the system architecture and the KG-first design. Section~\ref{sec:implementation} describes the implementation stack and data unification strategy. Section~\ref{sec:experiments} presents the experimental evaluation and results. Finally, Section~\ref{sec:conclusion} concludes with a discussion of limitations and future directions.

\section{Related Work}
\label{sec:related}

The design of intelligent competency services sits at the intersection of Semantic Web interoperability, neural information retrieval, and controlled generative AI. This section reviews the state of the art in these domains and identifies the specific operational gaps that \emph{SkillGraph-Service} aims to address.

Standardised vocabularies such as ESCO, ROME, and O*NET are the backbone of digital labour markets~\cite{luyen2025automated, le2025well}. These frameworks are typically published using the Simple Knowledge Organization System (SKOS) and accessed via SPARQL endpoints~\cite{miles2009skos, sparrql11}. While standards like IEEE Shareable Competency Definitions specify metadata structures for exchange~\cite{10194521}, they do not resolve the underlying semantic heterogeneity between frameworks.

Current alignment strategies often rely on either lexical matching, which fails on cross-lingual synonyms, or hard logic merging, which obscures the provenance of the original definitions. Furthermore, raw SPARQL access presents a high technical barrier for educational platform developers. There is a need for intermediate architectures that maintain the provenance of disparate frameworks (avoiding destructive merges) while exposing them through simplified, developer-friendly APIs.



Retrieval for skill discovery has evolved from purely lexical methods to complex hybrid architectures. The probabilistic BM25 model remains the robust baseline for exact terminology matching and short queries~\cite{robertson2009probabilistic}. However, BM25 struggles with the ``vocabulary mismatch'' inherent in educator queries (e.g., describing a skill without knowing its official title). To address this, dense retrieval utilizes bi-encoder architectures, such as Sentence-BERT~\cite{reimers2019sentence}, to map queries and documents into a shared semantic vector space. To scale these semantic lookups to large corpora, \textit{Approximate Nearest Neighbor} algorithms are required; \textit{Hierarchical Navigable Small World (HNSW)} graphs~\cite{malkov2018efficient} have emerged as a standard for balancing efficiency and recall.

Recent benchmarks demonstrate that fusing sparse (lexical) and dense (semantic) signals yields superior performance. Foundational strategies for combining these signals range from the linear score combination proposed by Fox and Shaw~\cite{fox1994combination} to rank-based methods like \textit{Reciprocal Rank Fusion}~\cite{cormack2009reciprocal}.
To further maximize precision, modern pipelines often employ a ``retrieve-then-rerank'' architecture. A second-stage Cross-Encoder -- which processes the query and document simultaneously -- typically provides the highest relevance scoring~\cite{nogueira2019passage}. However, in latency-sensitive microservices, the computational cost of cross-encoders is often prohibitive.


The application of LLMs to knowledge retrieval -- Retrieval-Augmented Generation (RAG) -- offers natural language interfaces for complex data~\cite{lewis2020retrieval}. However, hallucination remains a critical barrier in educational settings where factual accuracy is paramount. While techniques like prompting with knowledg-graph context improve controllability~\cite{liu2020k}, free-form generation often fails to strictly adhere to citation constraints.
Emerging research highlights the necessity of \emph{constrained decoding} (e.g., enforcing JSON schemas) to ensure that generative outputs remain auditable. There is a gap in the literature regarding the quantitative evaluation of such constraints specifically for competency explanations, comparing the trade-off between the fluency of free text and the faithfulness of structured outputs.

Many existing systems operate in silos: they either browse a single framework, focus solely on vector search, or rely on unconstrained LLM generation. \emph{SkillGraph-Service} distinguishes itself by integrating these concerns into a unified architecture: (i) Unlike hard-merge approaches, we implement a \textit{provenance-preserving KG layer} that unifies ESCO and ROME without data loss; (ii) Unlike heavy neural retrieval pipelines such as learned sparse models (SPLADE~\cite{formal2021splade}) and late-interaction dense retrievers (ColBERT~\cite{khattab2020colbert}), we propose and evaluate a \textit{lightweight hybrid stack} (SQLite+HNSW) that balances effectiveness with operational simplicity; (iii) Unlike standard RAG, we employ LLMs strictly for \textit{constrained ranking and schema-validated explanation}, prioritising auditability over creativity.

\section{Architecture Design}
\label{sec:architecture}

This section details the architecture and operational workflow of \emph{SkillGraph-Service}, a knowledge–graph-first microservice with LLM-enhanced capabilities for competence and skill retrieval. The system exposes a unified REST API consumed by heterogeneous learning and competence-management platforms (Moodle, Ikigai.games, FUN-MOOC, MEMORAe), and leverages a competency ontology built from labour-market and education standards (ROME, ESCO, O*NET). Figure~\ref{fig:system} summarises components and data flows.

\subsection{High-Level Overview}
The architecture is organised into four layers:

\begin{enumerate}
	\item \textit{API Surface and Consumer Layer} where clients invoke \texttt{/search}, \texttt{/prerequisites}, \texttt{/subskills} and \texttt{/explain}.
	\item \textit{Retrieval and Reasoning Layer} implementing hybrid retrieval (lexical + dense) and closed-candidate graph inference, with \emph{optional} LLM services.
	\item \textit{Knowledge and Representation Layer} hosting the Competency Management Ontology (CMO) plus aligned lexical/vector indices.
	\item \textit{Reference Standards Layer} containing ESCO and ROME, integrated into the CMO with provenance preservation. The design is extensible to additional frameworks (e.g., O*NET), but our current implementation and experiments focus on these two.
\end{enumerate}

All requests are handled in a KG-first, LLM-fallback manner: structured knowledge is exploited first; neural or generative components are used only for ranking, light inference, or audience-aware explanations  under constraints.

\subsection{Reference Standards Layer}
We ingest ROME/ESCO/O*NET exports (skills, occupations, multilingual labels, definitions, relations) into a common RDF/SKOS representation. Each entity carries:
\emph{stable identifiers}, \emph{hierarchical and associative relations} (\texttt{broader}, \texttt{narrower}, \texttt{related}, \texttt{prerequisite} when available), \emph{occupation links}, and \emph{provenance} (source, version, URI). Cross-framework links  based on an alignment between them are kept as mappings rather than hard merges to ensure reversibility and auditability.

\subsection{Knowledge and Representation Layer}
Building on the CMO introduced by Le~\cite{le2025vers}, the ontology supplies a unified schema that harmonises \emph{Skill}, \emph{Competence}, \emph{LearningOutcome}, and \emph{Occupation}. It captures core pedagogical and labour-market structure via relations such as \texttt{hasPrerequisite}, \texttt{hasSubSkill}, \texttt{isRelevantForOccupation}, and \texttt{isAssessedBy}, while modelling contextual facets (e.g., domain and level/difficulty) that later inform retrieval and explanation. The CMO is implemented in RDF/OWL and validated with SHACL constraints (e.g., every \texttt{Skill} must be linked to at least one framework or occupation). The resulting knowledge graph is hosted in an RDF store.

\subsection{Retrieval and Reasoning Layer}
\par{Competence and Skill Vectorisation:}
In parallel to the symbolic representation, we build dense vector representations for skills, competences and occupations. For each node in the CMO graph, we construct a textual representation by concatenating labels, descriptions, and optionally example usages. These texts are then encoded with a pre-trained sentence encoder (e.g., a Transformer-based model) to obtain fixed-size embeddings.
The embeddings are stored in a dedicated \emph{vector index}, which supports approximate nearest-neighbour search. This index is used for semantic similarity search, paraphrase matching, and cross-lingual retrieval between user queries and CMO concepts.

\par{Lexical and Symbolic Alignment:}
Each vector in the Lexical Index is linked back to its originating node in the CMO through a stable identifier. This alignment ensures that any semantic match obtained in the embedding space can be \emph{grounded} in the symbolic graph, enabling downstream reasoning such as prerequisite retrieval, sub-skill traversal, and explanation.

\par{Hybrid Retrieval:}
Educator queries are often short, written in different languages, and use varied wording for the same idea. To cope with this, the system searches in two complementary ways at once. First, it looks for exact and near-exact matches in the skill titles and descriptions, which keeps results precise when users employ standard terms. Second, it uses semantic similarity to find skills that “mean the same thing” even if the words are different or translated (e.g., FR/EN). The service then blends both views into a single ranking so that users see results that are both terminologically accurate and robust to paraphrases. When enabled, an optional lightweight re-ranking step refines the ordering of the top candidates while preserving low response times.

\par{Graph Retrieval:}
Beyond simple search, the service also reasons over the knowledge graph. Given a target skill, it first gathers its immediate neighbourhood: known prerequisites, declared sub-skills, related skills, and links to occupations. Because some frameworks do not list every prerequisite or sub-skill, the system builds a small, relevant ``candidate list'' around the target skill (close neighbours in the graph and skills tied to the same jobs). It then orders this list using a few transparent signals -- how close items are in the graph, how similar their names and descriptions are, and how often they connect to the same things. If evidence is weak and no explicit links exist, an LLM may help re-order only this small candidate list; it never invents new skills. When formal sub-skills are missing, the system can still suggest likely finer-grained skills from the neighbourhood and clearly label them as suggestions.

\par{LLM API Services}
The LLM component adds natural, audience-aware wording on top of the knowledge graph without replacing it. It is only used in a few controlled situations: (i) to write short explanations of a skill that cite the exact evidence used, (ii) to help order a small set of already-found candidates when explicit links are missing, and (iii) to suggest possible finer-grained skills drawn from a clearly defined neighbourhood. In all cases, the model works from facts the service provides (IDs, short snippets) and follows strict rules: produce exactly three concise sentences, attach 1–2 evidence IDs to each sentence, and stay in the requested language (FR/EN) and tone (teacher/learner). If the output does not respect these constraints, the system falls back to a deterministic template. The LLM never introduces skills that are not already in the candidate list, and prompts are kept minimal to protect privacy and keep latency predictable.

\begin{figure}[t]
	
	\centering\includegraphics[width=0.34\textwidth]{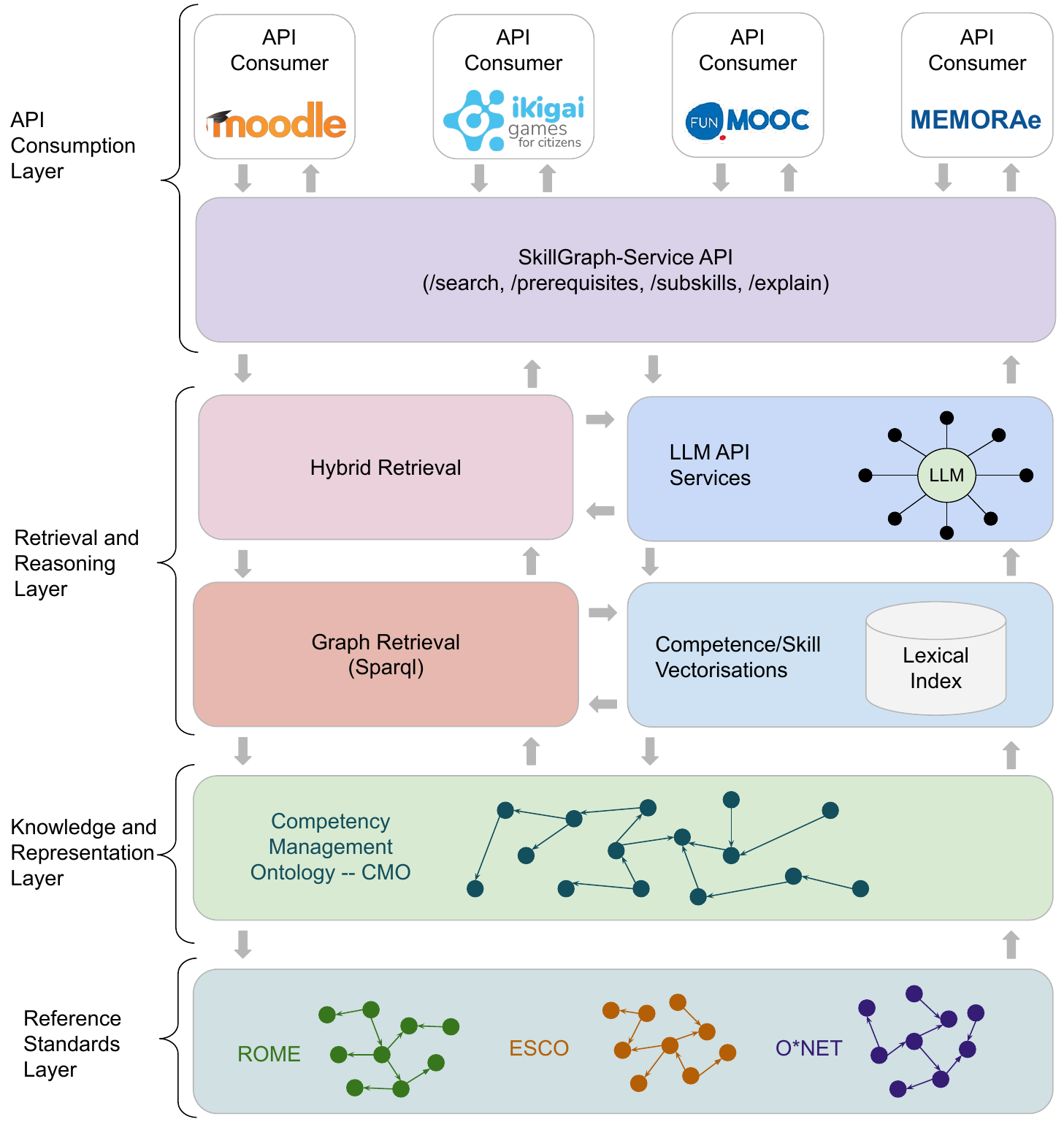}
	\caption{\emph{SkillGraph-Service} architecture: KG-first, hybrid retrieval, LLM-supported, minimal REST API.}
	\label{fig:system}
	\vspace{-0.7cm}
\end{figure}

\subsection{API Surface and Consumers Layer}
The service exposes a small, predictable REST surface that covers search, structure, granularity, and explanations. Client applications issue simple GET requests -- \texttt{/search?q=...,lang=fr,k=5} to retrieve top skills in a given language, \texttt{/skill/\{id\}} to fetch a canonical record, \texttt{/prerequisites?id,k} and \texttt{/subskills?id,k} to explore a skill’s neighbourhood, and \texttt{/explain?id,lang,audience} (optionally \texttt{format=\{text|json\}}) to receive a short, audience-aware explanation. Each response returns stable identifiers and labels, along with scores and source tags that indicate how results were obtained (\texttt{explicit}, \texttt{inferred-graph}, or \texttt{llm-candidate}); when explanations are in JSON, each sentence also carries the evidence IDs it cites.

A broad range of platforms can consume this API with minimal integration effort. Learning management systems (e.g., Moodle, FUN-MOOC\footnote{\url{https://www.fun-mooc.fr/fr/}}) map course outcomes to standardised skills and check prerequisites; serious-games platforms (e.g., Ikigai.games\footnote{\url{https://ikigai.games/}}) tag player actions with skills and suggest next steps;knowledge-management and collaborative learning platform (e.g., MEMORAe~\cite{abel2022memorae}) that supports learning innovation and visualises personal or organisational skill graphs. Because all endpoints rely on stable IDs and a handful of parameters, clients remain simple while benefiting from the full KG-first, hybrid reasoning pipeline behind the scenes.

\section{Implementation and Evaluation Methodology}
\label{sec:implementation}

The prototype operationalises the architecture proposed in Section~\ref{sec:architecture} via a lightweight, single-process stack designed to minimise operational overhead while maintaining low latency and deployment simplicity. The service is implemented in Python~3.11 using the FastAPI framework. To eliminate the complexity of external search clusters (e.g., Elasticsearch), the system integrates embedded retrieval engines -- \emph{SQLite FTS5}~\cite{sqlitefts5} for lexical operations and \emph{hnswlib}~\cite{hnswlib} for dense vector search -- directly within the application process.

\subsection{Data Ingestion and Unification}
\label{subsec:data-unification}
We integrate two primary European frameworks, ESCO and ROME, into a unified Knowledge Graph (KG). As summarised in Table~\ref{tab:framework-overview}, the ingested snapshot comprises \textit{14,257} ESCO skill nodes (multilingual across 28 locales) and \textit{10,319} ROME skill nodes (predominantly French).

\begin{table}[ht]
	\centering
	\caption{Dataset overview: Skill nodes, language coverage, and triple count per framework.}
	\label{tab:framework-overview}
	\begin{tabular}{l r r r}
		\toprule
		\textbf{Framework} & \textbf{Skill Nodes} & \textbf{Locales} & \textbf{Triples} \\
		\midrule
		ESCO & 14,257 & 28 (EN, FR, DE, \dots) & 1,493,678 \\
		ROME & 10,319 & 1 (FR) & 388,038 \\
		\bottomrule
	\end{tabular}
	\vspace{-0.2cm}
\end{table}

Data ingestion follows a strict provenance-preserving protocol. Source entities are ingested as RDF/SKOS triples with stable, namespaced identifiers. To prevent semantic drift, cross-framework correspondences are modelled as explicit mapping edges rather than destructive merges. Textual attributes (labels, descriptions) are Unicode-normalised and stored with language tags to support locale-specific retrieval.

\subsection{Lexical Retrieval}
\label{subsec:fts5}
Lexical search is handled by \textit{SQLite FTS5}~\cite{sqlitefts5}, configured with language-specific virtual tables containing \texttt{label}, \texttt{alt\_labels}, and \texttt{description} fields. Field weights are statically assigned ($W_{label} > W_{desc}$) to prioritise terminological precision.
Tokenisation strategies are language-dependent: English utilises Porter stemming, while French retains diacritics and disables stemming to prevent over-conflation of short acronyms common in French administrative nomenclature. Query execution prioritises phrase and prefix matching, falling back to token-based OR queries only when input syntax (e.g., special characters) precludes strict matching.

\subsection{Semantic Retrieval}
\label{subsec:hnswlib}
To address the vocabulary mismatch problem -- where educators use different terminology or languages to describe identical concepts -- we employ dense retrieval. Skill text representations are encoded into 768-dimensional vectors using a pre-trained multilingual Transformer model.
These embeddings are indexed using \textit{hnswlib}, an efficient implementation of Hierarchical Navigable Small World graphs~\cite{malkov2018efficient,hnswlib}. The index is held in-memory, enabling nearest-neighbour lookups (cosine similarity) with sub-millisecond latency. This architecture is designed to scale to hundreds of thousands of vectors without distributed infrastructure.

\subsection{Hybrid Fusion and Re-ranking}
\label{subsec:fusion}
The system combines lexical and semantic signals via a convex combination of normalised scores. Let $S_{lex}$ and $S_{sem}$ be the raw scores from FTS5 and hnswlib, respectively. The final score $S_{final}$ for a candidate skill $d$ is computed as:
\begin{equation}
	S_{final}(d) = \alpha \cdot \text{norm}(S_{lex}(d)) + (1 - \alpha) \cdot \text{norm}(S_{sem}(d))
\end{equation}
where $\text{norm}(\cdot)$ represents min-max normalisation over the retrieved candidate set, and $\alpha$ is a hyperparameter tuned on a held-out development set (default $\alpha=0.5$).

We evaluate four retrieval configurations: \textit{BM25-only} (Lexical baseline);
	\textit{Dense-only} (Semantic baseline); \textit{Hybrid} (Fused scoring); and \textit{Re-rank}, where the top-$N$ candidates are re-ordered using a cross-encoder or constrained LLM.

\subsection{Explanation Generation Module}
\label{subsec:explanations}
We implement three controlled generation modes to assess the trade-off between faithfulness and readability:
 \textit{C1 (Template):} A deterministic baseline that concatenates the definition, top-3 relations, and provenance. It serves as the high-speed fallback; \textit{C2 (Constrained LLM):} A generative approach constrained to produce a strict JSON schema comprising exactly three sentences. To enforce faithfulness, the model must cite specific evidence IDs (from the retrieved context) for every assertion;
 \textit{C3 (Free-form LLM):} An unconstrained generation mode used solely to benchmark readability, as it lacks enforceable citation mechanisms.

\subsection{Evaluation Metrics}
\label{subsec:metrics}
The evaluation is aligned with the two core research questions: (\textit{RQ1}) We evaluate retrieval performance using a test set of educator-formulated queries in French and English. Effectiveness is measured via \textit{Precision@k (P@k), k=\{3,5\}} and \textit{Normalised Discounted Cumulative Gain (nDCG@5)}. Efficiency is reported via median and 95th percentile (p95) latency per query;
(\textit{RQ2}) We assess explanation quality along three dimensions: \textit{Faithfulness (Hallucination Rate):} Measured via \emph{ID-based Coverage}, defined as the proportion of generated sentences containing valid evidence IDs found in the source graph context. For C2, we also calculate the Precision/Recall of these citations; \textit{Clarity:} quantified by average sentence length (readability proxy); \textit{Efficiency:} Measured by end-to-end generation latency.

\vspace{-0.2cm}
\section{Experiments and Results}
\label{sec:experiments}
\vspace{-0.2cm}
This section presents the empirical evaluation of the \emph{SkillGraph-Service}. We address the two primary research questions: \textit{RQ1}, concerning the trade-off between retrieval effectiveness and computational latency; and \textit{RQ2}, concerning the faithfulness and efficiency of generated explanations. 

\vspace{-0.2cm}
\subsection{RQ1: Search Effectiveness and Efficiency}
\label{subsec:rq1-results}

We evaluated retrieval performance using a curated set of 100 educator-oriented queries ($n_{\text{EN}}{=}50$, $n_{\text{FR}}{=}50$). These queries are characterised by domain-specific terminology and varying degrees of specificity.
Table~\ref{tab:retrieval} details the performance metrics (Precision@3, Precision@5, and nDCG@5) across four system configurations.

\begin{itemize}
	\item \textit{Baseline Performance:} The lexical baseline (\emph{BM25-only}) exhibits substantially lower ranking effectiveness (nDCG@5 $=0.474$), highlighting the vocabulary mismatch problem inherent in querying cross-jurisdictional skill standards.
	
	\item \textit{Impact of Dense Retrieval:} The \emph{Dense-only} variant yields a substantial improvement, improving nDCG@5 by $+84.4\%$ relative to the baseline. This confirms that semantic embedding spaces successfully capture the latent relationships between user queries and standardised skill descriptors.
	
	\item \textit{Hybrid Fusion:} The \emph{Hybrid} approach achieves the highest global effectiveness (P@5 $= 0.958$, nDCG@5 $= 0.948$). While the margin over \emph{Dense-only} is narrow, the inclusion of lexical signals proves critical for distinguishing between semantically similar but distinct concepts (e.g., ``Java programming'' vs.\ ``JavaScript programming''), particularly in English (nDCG@5 $= 0.970$).
	
	\item \textit{Re-ranking Overhead:} Unexpectedly, the \emph{Re-rank} configuration resulted in a degradation of nDCG@5 ($0.814$). We attribute this to the cross-encoder altering the ranking of the highly relevant top-$N$ candidates provided by the hybrid stage, potentially due to domain shift between the re-ranker's training data and the CMO.
\end{itemize}

\begin{table}[t]
	\centering
	\caption{Retrieval effectiveness ($n{=}100$ queries). Best results are highlighted in \textbf{bold}.}
	\label{tab:retrieval}
	\begin{tabular}{lccccc}
		\toprule
		\textbf{Variant} & \textbf{Lang} & \textbf{P@3} & \textbf{P@5} & \textbf{nDCG@5} \\
		\midrule
		\multirow{3}{*}{\emph{BM25-only}} 
		& EN & 0.460 & 0.340 & 0.600 \\
		& FR & 0.260 & 0.224 & 0.349 \\
		& \textbf{Avg} & 0.360 & 0.282 & 0.474 \\
		\midrule
		\multirow{3}{*}{\emph{Dense-only}} 
		& EN & \textbf{0.980} & \textbf{0.980} & 0.862 \\
		& FR & \textbf{0.940} & 0.932 & 0.887 \\
		& \textbf{Avg} & \textbf{0.960} & 0.956 & 0.875 \\
		\midrule
		\multirow{3}{*}{\textbf{\emph{Hybrid}}} 
		& EN & \textbf{0.980} & \textbf{0.980} & \textbf{0.970} \\
		& FR & 0.933 & \textbf{0.936} & \textbf{0.926} \\
		& \textbf{Avg} & 0.957 & \textbf{0.958} & \textbf{0.948} \\
		\midrule
		\multirow{3}{*}{\emph{Re-rank}} 
		& EN & \textbf{0.980} & 0.976 & 0.880 \\
		& FR & 0.920 & 0.924 & 0.748 \\
		& \textbf{Avg} & 0.950 & 0.950 & 0.814 \\
		\bottomrule
	\end{tabular}
	\vspace{-0.3cm}
\end{table}

Figure~\ref{fig:latency-bars} illustrates the latency distribution. The \emph{BM25-only} approach is negligible in cost (median 20\,ms). The \emph{Hybrid} system remains highly responsive, making it suitable for interactive autocomplete (median 187.5\,ms). However, the \emph{Re-rank} stage introduces a bottleneck, increasing median latency by an order of magnitude ($>2$\,s). Given the lack of effectiveness gain, the re-ranking step in its current configuration fails the cost-benefit analysis for real-time applications.

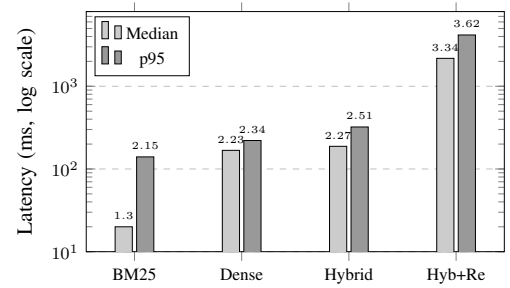
\begin{figure}[t]
	\centering
	\resizebox{0.75\columnwidth}{!}{
	\begin{tikzpicture}
		\begin{axis}[
			ybar,
			width=0.95\columnwidth,
			height=5.5cm,
			bar width=8pt,
			symbolic x coords={BM25,Dense,Hybrid,Hyb+Re},
			xtick=data,
			ymode=log, log basis y=10,
			ymin=10, ymax=8000,
			ylabel={Latency (ms, log scale)},
			ymajorgrids,
			grid style=dashed,
			legend style={at={(0.02,0.98)},anchor=north west,legend columns=1,font=\footnotesize},
			tick label style={font=\footnotesize},
			nodes near coords,
			nodes near coords style={font=\tiny, /pgf/number format/fixed},
			enlarge x limits=0.15
			]
			\addplot[fill=gray!40, draw=black] coordinates {
				(BM25,20) (Dense,168) (Hybrid,188) (Hyb+Re,2176)
			};
			\addplot[fill=gray!80, draw=black] coordinates {
				(BM25,140) (Dense,221) (Hybrid,322) (Hyb+Re,4170)
			};
			\legend{Median, p95}
		\end{axis}
	\end{tikzpicture}}
	\caption{Latency profile ($n{=}100$). Note the log scale. \emph{Hybrid} maintains interactive latencies ($<$200\,ms), while \emph{Re-ranking} incurs a significant penalty.}
	\label{fig:latency-bars}
	\vspace{-0.8cm}
\end{figure}

\subsection{RQ2: Explanation Fidelity and Quality}
\label{subsec:rq2-results}


We assess three explanation strategies: Template (C1), Constrained-JSON LLM (C2), and Free-form LLM (C3). We focus on the tension between \emph{readability} (natural language flow) and \emph{faithfulness} (accurate citation of graph evidence). Both LLM-based conditions (C2 and C3) are instantiated with the same open-weight Mixture-of-Experts model \texttt{gpt-oss-20b}~\cite{agarwal2025gpt}, so that differences in behaviour can be attributed to the decoding strategy (schema-constrained vs.\ free-form) rather than model capacity.

Results in Table~\ref{tab:explain-overall} reveal distinct operational profiles: \textit{C1:} Serves as the control for faithfulness. It achieves perfect coverage (1.0) and near-instant latency ($<5$\,ms). However, the output is rigid and structurally repetitive, reflected in the low word count per sentence; \textit{C2:} Offers a middle ground. The citation precision is perfect (1.0), indicating that when the model cites evidence, it does so correctly. However, the \emph{coverage} drops to 0.32, meaning the model frequently omits evidence IDs required by the schema, effectively ``forgetting'' to cite the source of its assertions. Despite this, it produces more verbose and natural text (22.5 words/sentence). The latency cost is high (median $\sim$2.2\,s); \textit{C3:} Without schema constraints, the model exhibits the highest unsupported rate (0.82). This confirms that unconstrained LLMs are unsuitable for verifiable ontology explanations where provenance is critical.

As shown in Table~\ref{tab:explain-overall}, the constrained model (C2) performs better in English than French regarding citation recall (0.83 vs 0.71, not shown in table). We hypothesise this stems from the model's stronger instruction-following capabilities in English, whereas French generation requires more tokens, increasing both latency and the probability of schema validation errors.

\begin{table}[t]
	\centering
	\caption{Explanation metrics. \emph{Cover.}: Coverage - fraction of sentences with valid evidence. \emph{Cit.\ P}: Precision of returned IDs against gold set. \emph{Lat}: Median latency.}
	\label{tab:explain-overall}
	\begin{tabular}{l c c c c r}
		\toprule
		\textbf{Mode} & \textbf{Cover.} & \textbf{Unsup.} & \textbf{Cit.\ P} & \textbf{W/Sent} & \textbf{Lat (ms)} \\
		\midrule
		C1 (Template)  & 1.00 & 0.00 & 1.00 & 5.1  & 4 \\
		C2 (JSON-LLM)  & 0.32 & 0.68 & 1.00 & 22.5 & 2221 \\
		C3 (Free-form) & 0.18 & 0.82 & 0.00 & 20.0 & 1622 \\
		\bottomrule
	\end{tabular}
	\vspace{-0.8cm}
\end{table}


The experimental results support the adoption of the \emph{Hybrid} retrieval strategy. By fusing dense vector retrieval with sparse lexical matching, the system mitigates the weaknesses of individual modalities: it handles the vocabulary mismatch of educator queries (where dense retrieval excels) while preserving precision for specific technical nomenclature (where lexical matching excels). The computational cost of this fusion is marginal ($\approx$20\,ms over Dense-only), making it viable for high-throughput environments.

The explanation results highlight a fundamental tension. Constraining an LLM to output strict JSON with citations (C2) guarantees that citations are valid (Precision=1.0) but significantly hampers the model's ability to cover all provided facts (Coverage=0.32). Conversely, free-form generation (C3) is fluent but unverifiable. For production systems requiring high trust, a \emph{hybrid user interface} is recommended: strictly structured template data (C1) should be displayed by default for immediate verifiable insight, with an optional ``Summarise'' action (C2) invoked asynchronously to provide a readable overview, explicitly labelled with its generated nature.

\section{Conclusion}
\label{sec:conclusion}

This paper presented \emph{SkillGraph-Service}, a microservice architecture designed to bridge the gap between rigid competency standards and the flexible needs of digital learning platforms. By integrating a provenance-aware Competency Management Ontology with a hybrid retrieval engine and LLM-based reasoning, we demonstrated a system that is both effective and architecturally lightweight.
Our evaluation confirms that a \textit{Hybrid retrieval approach} (FTS5 + HNSW) delivers the best search effectiveness among the tested configurations (nDCG@5 $>0.94$) with sub-200\,ms latency, suggesting that complex re-ranking steps may be unnecessary in our current domain setting. Furthermore, our investigation into explanation generation reveals that while LLMs can enhance readability, they struggle to maintain comprehensive citation coverage under strict schema constraints. Consequently, we advocate for a \textit{KG-first, LLM-fallback} design pattern: prioritising deterministic symbolic retrieval for correctness and speed, and reserving generative models for specific augmentation tasks where latency and probabilistic output are acceptable.
Future work will focus on improving the cross-lingual alignment of the vector space to boost French retrieval performance and fine-tuning smaller, local language models to reduce the latency of the explanation module.

\vspace{-0.1cm}		
\section*{Acknowledgments}\vspace{-0.1cm}	
We warmly thank the Ikigai consortium led by the association Games for Citizens, the company Gamaizer, as well as the FORTEIM project (winner of the AMI CMA France 2030 call for projects), for their support and collaboration. Their contributions have provided significant added value to the completion of this research.

\vspace{-0.1cm}		

	\bibliographystyle{IEEEtran}
	\bibliography{references} 
	
\end{document}